\documentclass{aastex}
\usepackage{emulateapj5}
\usepackage{apjfonts}
\input psfig.tex

\def\aj{{AJ}}

\def\apj{{ApJ}}
\def\apjs{{ApJS}}

\def\etal{{et al. }}

\def\gax{{$\mathrel{\hbox{\rlap{\hbox{\lower4pt\hbox{$\sim$}}}\hbox{$>$}}}$}}

\def\kms{km s$^{-1}$}

\def\lax{{$\mathrel{\hbox{\rlap{\hbox{\lower4pt\hbox{$\sim$}}}\hbox{$<$}}}$}}
\def\lum{erg s$^{-1}$}

\def\mnras{{MNRAS}}
\def\nat{{Nature}}

\def\percm2{cm$^{-2}$}

\def\solmass{$M_\odot$}

\received{}
\accepted{}
\slugcomment{To appear in {\it The Astrophysical Journal (Letters).}}
\shorttitle{ORIGIN OF RADIO EMISSION IN LLAGNS}
\shortauthors{ULVESTAD \& HO}
\journalid{}{}
\articleid{}{}

\begin{document}

\title{The Origin of Radio Emission in Low-Luminosity Active Galactic Nuclei: 
Jets, Accretion Flows, or Both?}

\author{James S.~Ulvestad\altaffilmark{1} and Luis C.~Ho\altaffilmark{2}}

\altaffiltext{1}{National Radio Astronomy Observatory, P.O. Box O, Socorro, 
NM 87801; julvesta@nrao.edu.}

\altaffiltext{2}{The Observatories of the Carnegie Institution of Washington,
813 Santa Barbara St., Pasadena, CA 91101; lho@ociw.edu.}

\begin{abstract}

The low-luminosity active galactic nuclei in NGC~3147, NGC~4203, and NGC~4579 have been 
imaged at four frequencies with the Very Long Baseline Array.  The galaxies are 
unresolved at all frequencies, with size upper limits of $10^3$--$10^4$ 
times the Schwarzschild radii of their central massive black holes.  The 
spectral indices between 1.7 and 5.0~GHz 
range from 0.2 to 0.4; one and possibly two of the galaxies show spectral 
turnovers between 5.0 and 8.4~GHz.  The high brightness temperatures 
($T_b$\gax $10^9$~K) and relatively straight spectra imply that free-free emission 
and/or absorption cannot account for the slightly inverted spectra.  Although 
the radio properties of the cores superficially resemble predictions for
advection-dominated accretion flows, the radio luminosities are too high 
compared to the X-ray luminosities.  We suggest that the bulk of the radio 
emission is generated by a compact radio jet, which may coexist with a low 
radiative efficiency accretion flow.

\end{abstract}


\section{Introduction}

Most nearby normal and active galaxies with bulges appear to have supermassive 
black holes (BHs) at their centers (see Kormendy \& Gebhardt 2001 for a review), 
although galaxies without bulges may lack BHs (Gebhardt et al. 2001).  The 
central BHs are thought to power the low-luminosity active galactic nuclei 
(LLAGNs) in local Seyfert galaxies and LINERs.  In a number of these LLAGNs, 
the radio emission is compact on milliarcsecond (mas) scales (Falcke et al. 
2000; Nagar, Wilson, \& Falcke 2001) 
when imaged with the Very Long Baseline Array (VLBA; Napier et al. 
1993)\footnote{The VLBA is operated by the National Radio Astronomy 
Observatory, a facility of the National Science Foundation operated under 
cooperative agreement by Associated Universities, Inc.}.  They often have 
mildly inverted centimeter radio spectra with spectral indices $\alpha \approx 
+0.3$, where $S_\nu \propto \nu^{\alpha}$ (Wrobel \& Heeschen 1991; 
Slee et al. 1994; Nagar et al. 
2000). In our recent VLA study of the Palomar Seyfert sample 
(Ho, Filippenko, \& Sargent 1997), we found that about half of the weak Seyferts 
selected by optical spectroscopy have flat or inverted radio spectra 
(Ho \& Ulvestad 2001, hereafter HU01; Ulvestad \& Ho 2001), 
These galaxies differ from more luminous Seyferts, which typically have optically 
thin synchrotron spectra with $\alpha\approx -0.7$ (Ulvestad \& Wilson 1984, 
1989; Ulvestad \& Ho 2001).  Possible explanations 
discussed by Ulvestad \& Ho (2001) included free-free emission/absorption or an 
advection-dominated accretion flow (ADAF; see Narayan, Mahadevan, \& Quataert 
1998; Quataert 2001).

The Galactic Center source Sgr~A* also has a mildly inverted spectrum at 
centimeter wavelengths, with $\alpha\approx +0.3$ from 1~GHz to over 100~GHz 
(Falcke et al. 1998).  This spectrum is sometimes attributed to an ADAF (e.g., 
Narayan, Yi, \& Mahadevan 1995; Manmoto, Mineshige, \& Kusunose 1997). In 
ADAFs, synchrotron radiation from hot electrons in a $\geq 10^9$~K gas is 
self-absorbed, yielding a radio spectral index between +0.3 and +0.4 
(Mahadevan 1997; Narayan et al. 1998).  X-ray emission arises from Compton 
scattering or bremsstrahlung, but is far below the 
Eddington limit, since most of the plasma's energy is advected into the BH by 
the ions.  The transition from a classical thin accretion disk to an ADAF takes 
place within $\sim10^3-10^4$ Schwarzschild radii ($R_{\rm S}$) of the BH.  
Compact radio emission from a number of LLAGNs has been attributed to ADAF 
processes similar to those inferred in Sgr~A* (e.g., Fabian \& Rees 1995; 
Mahadevan 1997; Ho et al. 2000; Wrobel \& Herrnstein 2000).  However, Di~Matteo 
et al. (1999) find that the simplest ADAF models overpredict the radio and 
submillimeter fluxes in several nearby ellipticals. 

Another possible model for Sgr~A* and LLAGNs is that of a compact radio jet 
in combination with an ADAF (Falcke 1996; Falcke \& Markoff 2000; Yuan 
2000).  As reviewed by Falcke (2001), the jet 
accelerates slightly due to a longitudinal pressure gradient, and 
integration of jet radio emission may result in a 
slightly inverted spectrum.  Compton upscattering of the radio photons produces 
the X-rays.  An energetically significant jet or 
outflow will reduce the accretion rate in the innermost region of the 
ADAF, which may explain the low level of radio emission 
seen in some ellipticals (Di~Matteo et al. 1999; Quataert \& Narayan 1999).

In this {\it Letter}, we report four-frequency VLBA imaging of 
three LLAGNs imaged previously at arcsecond resolution with the VLA (HU01).  These galaxies 
were selected because they have inverted VLA spectra between 1.4 and 4.9~GHz 
and flux densities sufficiently high (\gax 10~mJy) for easy imaging
with the VLBA.  Therefore, 
they were good candidates for testing the various models that might account 
for their spectral shapes.

\section{Observations and Results}

The VLBA observed NGC~3147, NGC~4203, and NGC~4579 
in March 2001 at central frequencies of 1.667, 2.271, 
4.995, and 8.421 GHz.  About six hours was spent integrating on each galaxy, divided 
among the four frequencies.  Amplitude calibration used a-priori gain
values together with system temperatures measured
during the observations, and typically is accurate to 5\%.  Initial clock and 
atmospheric (phase) errors were derived from calibrator sources within 
2\arcdeg --4\arcdeg\ of the galaxies, 

\begin{figure*}[t]
\centerline{\psfig{file=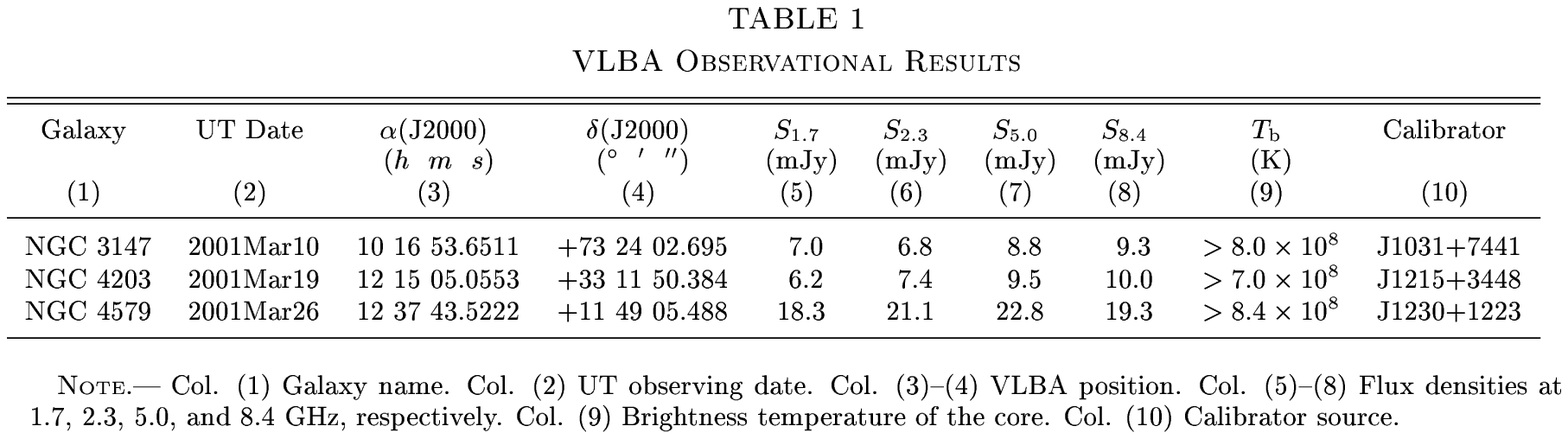,width=19.0cm,angle=0}}
\end{figure*}

\begin{figure*}[t]
\centerline{\psfig{file=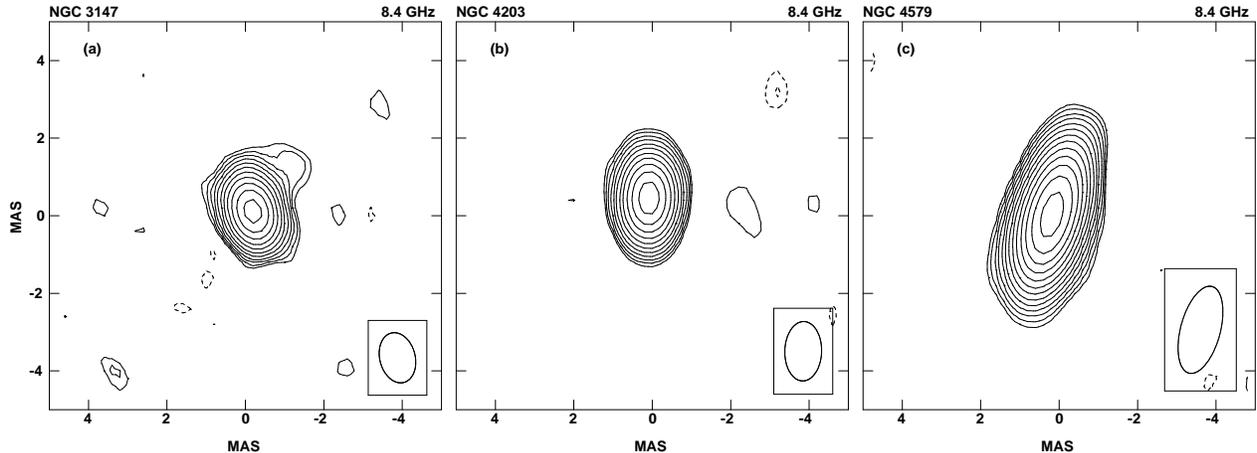,width=17.5cm,angle=0}}
\figcaption[fig1.ps]{
8.4~GHz VLBA images of ({\it a\/}) NGC~3147, ({\it b\/}) NGC~4203, and
({\it c\/}) NGC~4579.  All images are 9.6~mas on a side, with contour intervals
increasing by factors of $\sqrt{2}$ from 0.25~mJy~beam$^{-1}$ (negative
contours are shown dashed).  The restoring beam for each image is shown in
the lower-right corner.
\label{fig1}}
\end{figure*}
\vskip 0.3cm
 
\noindent
using the technique of phase-referencing 
(Beasley \& Conway 1995).  This resulted in galaxy position errors of
\lax~1~mas, dominated by uncertainties in the calibrator positions 
(from Johnston et al. 1995; A.~J.~Beasley et al., in preparation).
The initial phase calibration was limited at low frequencies by 
the active ionosphere, and at high frequencies by
the wet troposphere, but each galaxy was detected at all four 
frequencies.  Phase self-calibration was 
applied iteratively, resulting in noise-limited images.
Approximate beam widths ranged from $\sim 7$~mas at 1.7 GHz
to 1.2~mas at 8.4~GHz, while the rms image noises were
90--150~$\mu$Jy~beam$^{-1}$.

Gaussian fitting showed that all VLBA images were dominated by a single 
unresolved component; we estimate size upper limits of half the 
beam widths.  Derived flux densities are 
accurate to $\sim 5$\% at 2.3 and 5.0~GHz, and $\sim 10$\% (due to 
larger self-calibration corrections) at 1.7 and 8.4~GHz.  Table~1 summarizes 
the observations and main results, and Table~2 lists some derived quantities 
and other pertinent information.  Figure~1 shows the high-resolution, 
unresolved 8.4-GHz images of the galaxies, while Figure~2 compares the radio 
spectra with that of Sgr~A*. 

\section{Individual Galaxies}

{\it NGC~3147}.---NGC~3147 contains a Seyfert 2 nucleus in an Sbc galaxy
(Ho et al.~1997).  The hard X-ray (2--10 keV) luminosity 
measured with {\it ASCA}\ is $3.4\times 10^{41}$erg~s$^{-1}$ (Terashima, Ho, \& 
Ptak 2000a), and observations with the {\it ROSAT}\ \ HRI (Roberts \& Warwick 
2000) show that this source is almost certainly located at the nucleus.  Our 
upper limit to the source 
radius is $\sim$0.05 pc, or $\sim 1500 R_{\rm S}$ for a BH 
mass of $3.6\times10^8$ \solmass.  We estimated the BH mass using the tight 
empirical correlation between BH mass and bulge stellar velocity dispersion 
(Gebhardt et al. 2000; Ferrarese \& Merritt 2000), adopting the 
$M_{\rm BH}-\sigma$ relation given by Gebhardt et al. (2000) and $\sigma$ = 
268 \kms\ from McElroy (1995)\footnote{Gebhardt et al. (2000) use $\sigma = 
\sigma_e$, the projected, luminosity-weighted velocity dispersion measured 
within the effective radius of the bulge.  Since our objects do not have 
measurements of $\sigma_e$, we use $\sigma = \sigma_0$, the central velocity
dispersion; Gebhardt et al. (2000) show that $\sigma_e\,\approx\, \sigma_0$ 
within a scatter of $\sim$10\%.}.  The VLBA flux densities of 7 and 9~mJy at 1.7 
and 5.0 GHz can be compared to the VLA peak flux densities of 13 and 10~mJy at 
similar frequencies in late 1999 (HU01), so there may be weak
1.7-GHz emission on scales between tens of mas and 1~arcsec.

{\it NGC 4203}.---NGC~4203 is a nearly face-on S0 galaxy with a LINER~1.9 nucleus 
(Ho et al.~1997).  It has a double-peaked broad H$\alpha$ emission line with a 
full-width near zero intensity of at least 12,500 \kms\ (Shields et al. 2000), 
and a fairly weak nuclear X-ray source detected with {\it Chandra},
having $L_{\rm X}$(2--10~keV) = $5.0\times 10^{39}$~erg~s$^{-1}$ (Ho et al. 2001).
For $\sigma$ = 124 \kms\ (Dalle~Ore et al. 1991), $M_{\rm BH} = 2.0\times10^7$
\solmass.  The upper limit for the source radius is 
$\sim$0.02 pc, or $1.0\times 10^4 R_{\rm S}$, and the VLBA flux 
densities are consistent with the 1999 VLA peaks (HU01).

{\it NGC 4579}.---NGC~4579 is an SABb galaxy with a type~1.9 Seyfert or LINER 
nucleus (Ho et al.~1997).  Its broad H$\alpha$ line is double-peaked or
double-shouldered (Barth et al.  2001), and it contains a variable hard X-ray 
source (Terashima et al. 2000b) with a 2--10~keV luminosity of $\sim 8.9\times 
10^{40}$ \lum\ as measured with {\it Chandra}\ (Ho et al. 2001).  The BH mass 
inferred from the 

\begin{figure*}[t]
\centerline{\psfig{file=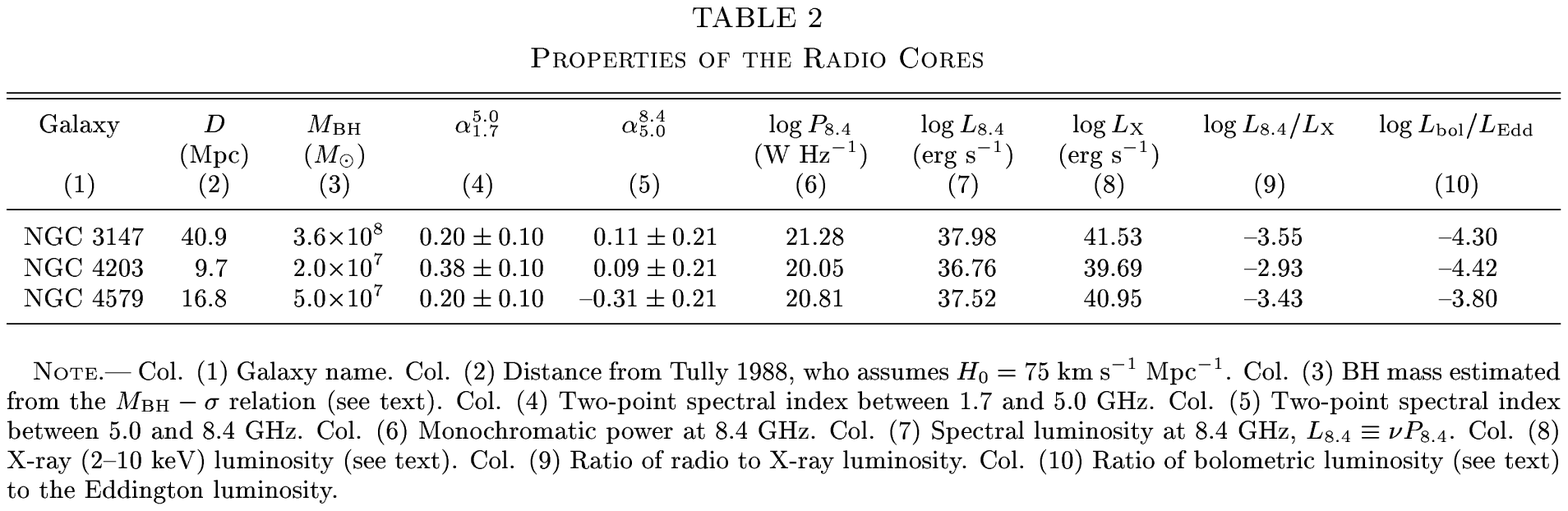,width=19.0cm,angle=0}}
\end{figure*}

\noindent
$M_{\rm BH}-\sigma$ relation is $\sim 5\times10^7$ \solmass\ 
(Barth et al. 2001).  Our VLBA 8.4-GHz beam of 2.4 mas $\times$ 1.0 
mas limits the source radius to $\sim$0.03 pc, or $\sim 6200 R_{\rm S}$.  The 
1.7-GHz VLBA flux density is consistent with 
the previous VLA peak  (HU01), while the 5-GHz value is 
considerably lower (22.8 mJy vs. 38~mJy).  Falcke et al. (2001) found this 
galaxy to be variable at 15~GHz on time scales of 1--3~yr, so the change in 
the 5-GHz flux density over 17 months is most likely due to variability.  
This variability implies that the core spectral index changes over time, 
so model interpretations based solely on the spectral index must be made with 
caution.

\section{Implications and Summary}

What is the emission mechanism of the radio cores? In the
Seyfert galaxy NGC~1068, Gallimore, Baum, \& O'Dea (1997) resolved
the flat-spectrum component S1 at
8.4~GHz using the VLBA; they attributed S1
to free-free emission from a parsec-scale torus. However, other flat-spectrum
Seyfert cores are unresolved by the VLBA and appear to have other origins
(Mundell et al.~2000).  The free-free interpretation also is 
untenable for our sources because (1) they are much more compact 
($r$ \lax\ 0.05~pc) than the dimensions of nuclear tori and (2) they have 
brightness temperatures ($T_b$ \gax\ $10^9$ K; Table~1) more than two orders of 
magnitude above the values expected for thermal emission.  The relatively flat 
spectra over a factor of 5 in frequency (Fig.~2) seem to rule out 
free-free absorption as well, though NGC~4579 shows a slight
spectral turnover above 5~GHz.  Therefore, the 
prevalence of flat or inverted spectra in LLAGNs cannot, in general, be due to 
free-free emission or absorption.  The brightness temperature limits are 
consistent with classical synchrotron self-absorption, although more than one 
self-absorbed component would have to be present, since self-absorbed 
synchrotron radiation from a single power-law distribution of electrons should
result in $\alpha \approx 2.5$. 

If our galaxies host massive BHs, as suggested by current BH demography studies 
and the evidence for nonstellar activity summarized in \S~3, they must be 
radiating significantly below their Eddington limits. The nuclear bolometric 
luminosities of NGC 4203 and NGC 4579 are known from Ho et al. (2000) and Ho 
(1999), respectively.  For NGC 3147, we assume $L_{\rm bol} = 
6.7 L_{\rm X}$(2--10~keV), empirically derived from an average of 10 objects 
analyzed by Ho (1999) and Ho et al. (2000).  As shown in Table~2, all three 
objects are highly sub-Eddington systems: $L_{\rm bol}/L_{\rm Edd}$ \lax\ 
$10^{-4}$.  In particular, they all lie comfortably within the expected value
for ADAFs, which is estimated to be 
$L_{\rm bol}/L_{\rm Edd}\,\lesssim\,10^{-2}$ (e.g., Narayan et al. 1998; 
Quataert 2001).  Additional arguments for the existence of ADAFs
in NGC 4203 and NGC 4579 are given in Ho et al. (2000) and Ho (2001), 
among them being the absence of the optical/ultraviolet continuum bump 
normally attributed to thermal emission from a thin disk.  Thus, the 
compact radio cores could be emission from ADAFs.  This is consistent with 
(1) the compactness of the sources ($r < 10^4 R_{\rm S}$), (2) the apparent 
self-absorbed synchrotron nature of the emission, (3) the brightness temperature 
limits, and (4) the spectral indices\footnote{The apparent spectral turnover in 
NGC 4579, however, would require some modification of the standard ADAF model, 
such as inclusion of winds (Quataert \& Narayan 1999).}.

Notwithstanding these arguments, the majority of the radio emission most likely 
does {\it not}\ come from accretion flows.  We arrive at this conclusion by comparing 
the luminosity output of the radio and the X-ray bands.  Table~2 shows that the 
ratio of the spectral luminosity at 8.4~GHz to the luminosity in the 2--10 keV 
band spans $3\times 10^{-4}$ \lax  $L_{8.4}/L_{\rm X}$ \lax  $1\times 10^{-3}$.  Yi \& 
Boughn (1998) give a convenient expression for the relation between radio 
and X-ray luminosities in a standard ADAF model.  Using our values of BH mass 
and X-ray luminosity in their Equation 2.10, we predict values for 
$L_{8.4}/L_{\rm X}$ that are $\sim$10 times smaller than the observed values.  
Hence our sources are overluminous in the radio compared to the X-rays.  We note 
that this discrepancy is not alleviated by appealing to more recent 
variants of the ADAF model that incorporate outflows or convection 
(e.g., Quataert 2001).  Either of these 
effects will suppress the radio component with respect to the X-rays 
(Quataert \& Narayan 1999; Ball, Narayan, \& Quataert 2001), 
exacerbating the problem.  If the radio core were larger than in the
canonical ADAF models, the surrounding X-ray bremsstrahlung region
would be correspondingly larger, again increasing the discrepancy 
between our high observed values of $L_{8.4}/L_{\rm X}$ and the 
lower model values.

The most likely origins for the radio cores in our three galaxies 
are compact jets or outflows, as discussed by Falcke (2001) and
Nagar et al.~(2001).  The high brightness 
temperatures, the flat/inverted spectra, and the compactness of the cores 
easily can be reproduced in such a model.  However, it is unclear 
if the jet model can self-consistently explain both the radio and 
X-ray emission, let alone the entire spectral energy distribution.  
Falcke \& Markoff (2000) successfully fitted the radio and X-ray spectra 
of Sgr~A$^*$ with the jet model, but $L_{8.4}/L_{\rm X}$ 
in Sgr~A$^*$ 

\vskip 0.3cm
\psfig{file=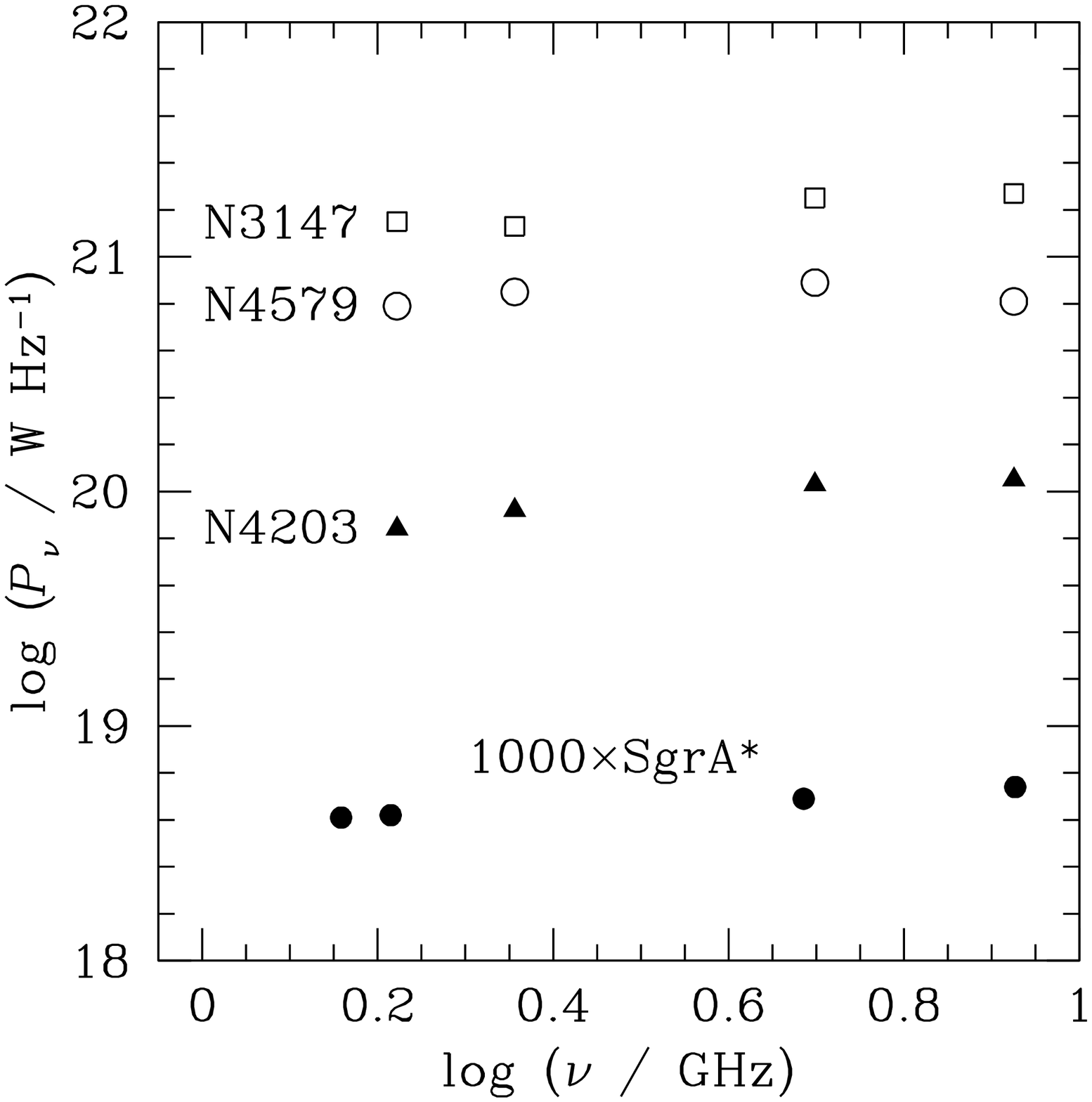,width=8.5cm,angle=0}
\figcaption[fig2.ps]{
VLBA radio spectra of the unresolved cores of the three LLAGNs.  The spectrum
of the Galactic Center source Sgr~A* (Falcke et al. 1998) is shown for
comparison, assuming a distance of 8.0~kpc (Reid 1993), with power scaled up
by a factor of 1000.  Error bars for the powers are smaller than or equal
to the symbol size.
\label{fig2}}
\vskip 0.3cm

\noindent
is \gax 100 times higher than in our objects during
the quiescent X-ray phase found by Baganoff et al. (2001).
For a given radio jet power, the X-ray luminosities from self-Compton emission
are well-constrained (e.g., Falcke \& Markoff 2000), and are unlikely to
contribute significantly in our galaxies.

A natural way to account simultaneously for the full spectral energy 
distribution of LLAGNs, such as those discussed here, is to incorporate
facets of both classes of models discussed above, namely a jet and an ADAF (or 
some closely related variant of a low radiative efficiency accretion flow; 
see Quataert 2001).  This is the approach taken by Yuan (2000), but we note 
that no existing model self-consistently incorporates both components.  

\acknowledgments

We thank Heino Falcke for helpful discussions and Nadia Afrin for assistance 
in generating the VLBA observing schedules.  This research has made use of 
NASA's Astrophysics Data System Abstract Service and the NASA/IPAC 
Extragalactic Database (NED) which is operated by the Jet Propulsion Laboratory, 
California Institute of Technology, under contract with the National Aeronautics 
and Space Administration.  The work of L.~C.~H. is partly funded through NASA grants 
awarded by the Space Telescope Science Institute, which is operated by AURA, 
Inc., under NASA contract NAS5-26555.



\begin{thebibliography}{}

\bibitem[]{} 
Baganoff, F.~K., et al. 2001, \nat, 413, 45

\bibitem[]{} 
Ball, G.~H., Narayan, R., \& Quataert, E. 2001, \apj, 552, 221

\bibitem[]{} 
Barth, A.~J., Ho, L.~C., Filippenko, A.~V., Rix, H.-W., \& Sargent, W.~L.~W. 
2001, \apj, 546, 205

\bibitem[]{}
Beasley, A. J., \& Conway, J. E. 1995, in Very Long Baseline Interferometry and 
the VLBA, ed. J.~A. Zensus, P.~J. Diamond, \& P.~J. Napier (San Francisco: 
ASP), 327

\bibitem[]{}
Dalle~Ore, C., Faber, S.~M., Jesus, J., Stoughton, R., \& Burstein, D. 1991,
\apj, 366, 38

\bibitem[]{}
Di~Matteo, T., Fabian, A.~C., Rees, M.~J., Carilli, C.~L., \& Ivison, R.~J.
1999, \mnras, 305, 492

\bibitem[]{}
Fabian, A.~C., \& Rees, M.~J. 1995, \mnras, 277, L55

\bibitem[]{}
Falcke, H. 1996, \apj, 464, L67

\bibitem[]{}
------. 2001, Reviews in Modern Astronomy, 14, in press

\bibitem[]{}
Falcke, H., Goss, W.~M., Matsuo, H., Teuben, P., Zhao, J.-H., \& Zylka, R.
1998, \apj, 499, 731

\bibitem[]{}
Falcke, H., Leh\'ar, J., Barvainis, R., Nagar, N.~M., \& Wilson, A.~S. 2001, 
in Probing the Physics of Active Galactic Nuclei by Multiwavelength Monitoring, 
ed. B.~M. Peterson, R.~S. Polidan, \& R.~W. Pogge (San Francisco: ASP), 265

\bibitem[]{}
Falcke, H., \& Markoff, S. 2000, \aap, 362, 113

\bibitem[]{}
Falcke, H., Nagar, N.~M., Wilson, A.~S., \& Ulvestad, J.~S. 2000, \apj, 542, 197

\bibitem[]{}
Ferrarese, L. \& Merritt, D. 2000, \apj, 539, L9

\bibitem[]{}
Gallimore, J.~F., Baum, S.~A., \& O'Dea, C.~P. 1997, \nat, 388, 852

\bibitem[]{}
Gebhardt, K., et al. 2000, \apj, 539, L13

\bibitem[]{}
------. 2001, \aj, 122, in press, 2001Nov (astro-ph/0107135)

\bibitem[]{}
Ho, L.~C. 1999, \apj, 516, 672

\bibitem[]{}
------. 2001, in Issues in Unification of AGNs, ed. R. Maiolino, A. Marconi,
\& N. Nagar (San Francisco: ASP), in press

\bibitem[]{}
Ho, L.~C., \etal 2001, \apj, 549, L51

\bibitem[]{}
Ho, L.~C., Filippenko, A.~V., \& Sargent, W.~L.~W. 1997, \apjs, 112, 315

\bibitem[]{}
Ho, L.~C., Rudnick, G., Rix, H.-W., Shields, J.~C., McIntosh, D.~H.,
Filippenko, A.~V., Sargent, W.~L.~W., \& Eracleous, M. 2000, \apj, 541, 120

\bibitem[]{}
Ho, L.~C., \& Ulvestad, J.~S. 2001, \apjs, 133, 77 (HU01)

\bibitem[]{}
Johnston, K.~J., et al. 1995, \aj, 110, 880

\bibitem[]{}
Kormendy, J., \& Gebhardt, K. 2001, in The 20th Texas Symposium on Relativistic
Astrophysics, ed. H. Martel \& J.~C. Wheeler (New York: AIP), in press 
(astro-ph/0105230)

\bibitem[]{}
Mahadevan, R. 1997, \apj, 477, 585

\bibitem[]{} 
Manmoto, T., Mineshige, S., \& Kusunose, M. 1997, \apj, 489, 791

\bibitem[]{} 
McElroy, D.~B. 1995, \apjs, 100, 105

\bibitem[]{} 
Mundell, C.~G., Wilson, A.~S., Ulvestad, J.~S., \& Roy, A.~L. 2000, \apj, 529, 816

\bibitem[]{}
Nagar, N.~M., Falcke, H., Wilson, A.~S., \& Ho, L.~C. 2000, \apj, 542, 186

\bibitem[]{}
Nagar, N.~M., Wilson, A.~S., \& Falcke, H.~2001, \apj, 559, L87

\bibitem[]{}
Napier, P.~J., Bagri, D.~S., Clark, B.~G., Rogers, A.~E.~E., Romney, J.~D., 
Thompson, A.~R., \& Walker, R.~C. 1993, Proc IEEE, 82, 658

\bibitem[]{}
Narayan, R., Mahadevan, R., \& Quataert, E. 1998, in The Theory of Black Hole
Accretion Discs, ed.  M. A. Abramowicz, G. Bj\"{o}rnsson, \& J. E. Pringle
(Cambridge: Cambridge Univ. Press), 148

\bibitem[]{}
Narayan, R., Yi I., \& Mahadevan, R. 1995, \nat, 374, 623


\bibitem[]{} 
Quataert, E. 2001, in Probing the Physics of Active Galactic Nuclei by
Multiwavelength Monitoring, ed. B.~M. Peterson, R.~S. Polidan, \& R.~W. Pogge
(San Francisco: ASP), 71

\bibitem[]{}
Quataert, E., \& Narayan, R. 1999, \apj, 520, 298

\bibitem[]{}
Reid, M. J. 1993, \araa, 31, 345

\bibitem[]{}
Roberts, T.~P., \& Warwick, R.~S.  2000, \mnras, 315, 98

\bibitem[]{}
Shields, J.~C., Rix, H.-W., McIntosh, D.~H., Ho, L.~C., Rudnick, G.,
Filippenko, A.~V., Sargent, W.~L.~W., \& Sarzi, M. 2000, \apj, 534, L27

\bibitem[]{}
Slee, O.~B., Sadler, E.~M., Reynolds, J.~E., \& Ekers, R.~D. 1994, \mnras, 269,
928

\bibitem[]{}
Terashima, Y., Ho, L.~C., \& Ptak, A.~F. 2000a, \apj, 539, 161

\bibitem[]{}
Terashima, Y., Ho, L.~C., Ptak, A.~F., Yaqoob, T., Kunieda, H., Misaki, K.,
\& Serlemitsos, P.~J. 2000b, \apj, 535, L79

\bibitem[]{}
Tully, R.~B. 1988, Nearby Galaxies Catalog (Cambridge: Cambridge Univ. Press)

\bibitem[]{}
Ulvestad, J. S., \& Ho, L. C. 2001, \apj, 558, 561 

\bibitem[]{}
Ulvestad, J.~S., \& Wilson, A.~S. 1984, \apj, 285, 439

\bibitem[]{}
------. 1989, \apj, 343, 659

\bibitem[]{}
Wrobel, J.~M., \& Heeschen, D. 1991, AJ, 101, 148

\bibitem[]{}
Wrobel, J.~M., \& Herrnstein, J.~R. 2000, \apj, 533, L111

\bibitem[]{}
Yi, I., \& Boughn, S.~P. 1998, \apj, 499, 198

\bibitem[]{}
Yuan, F. 2000, \mnras, 319, 1178

\end{thebibliography}
\end{document}